# Which differences can be expected when two universities in the Leiden Ranking are compared? Some benchmarks for institutional research evaluations

Lutz Bornmann* & Wolfgang Glänzel**

*Division for Science and Innovation Studies
Administrative Headquarters of the Max Planck Society
Hofgartenstr. 8,
80539 Munich, Germany.
Email: bornmann@gv.mpg.de

** ECOOM and Department of MSI
KU Leuven
Naamsestraat 61,
3000 Louvain, Belgium.

Department of Science Policy and Scientometrics
Library of the Hungarian Academy of Sciences
Arany János u. 1,
Budapest 1051, Hungary.
Email: wolfgang.glanzel@kuleuven.be; glanzw@iif.hu

**Abstract**
The comparison of two universities in terms of bibliometric indicators frequently faces the problem of assessing the differences as meaningful or not. This Letter to the Editor proposes some benchmarks which can be used for supporting the interpretation of institutional differences.

University rankings are frequently used to assess the performance of universities. The Leiden Ranking (see http://www.leidenranking.com) is one of the most popular international university rankings which is exclusively based on bibliometric indicators (e.g. number of publications and mean number of citations). The focus on bibliometrics avoids, e.g., one of the most critical points of other university rankings: the arbitrary use of weights to combine different indicators to an overall score. There are two important problems with the interpretation of the results of university rankings. The first problem concerns the interpretation of the ranking positions. Many universities have very different ranking positions, although the performance of the universities is very similar. Thus, Bornmann and Glänzel (2017) recommend categorizing the universities in four meaningful groups on the basis of one indicator, which reveals whether the performance is poor, fair, remarkable, or outstanding. This clustering leads (for most of the universities) to performance statements on universities which are not dependent on small performance differences. The second problem with the interpretation of university rankings concerns the interpretation of differences between two universities. In many cases, users of the Leiden Ranking are interested in pairwise performance differences. If confidence intervals for individual indicator values or indicator components are given, the formal significance of the individual deviation of two universities can be judged at this level (Bornmann, Mutz, & Daniel, 2013). More generally, the problem is, however, that it is difficult to assess for the user how meaningful the differences between the scores of two universities are. Is there a meaningful performance difference if the scores differ by 0.5, or does meaningfulness start only with a difference of 1.0?

In this Letter to the Editor, we present benchmarks which support the interpretation of performance differences. The benchmark calculations are based on the "Results" worksheet with the results of the Leiden Ranking 2017 from http://www.leidenranking.com/downloads. The worksheet contains results for 903 universities across seven time periods. The results are presented using full and fractional counting. In this study, we focus on full counting and "All sciences". However, the same analyses can in principle be performed with a focus on fractional counting and specific fields. We considered the following indicators in this study (see http://www.leidenranking.com/information/indicators):

MNCS: the average number of citations of the publications of a university, normalized for field and publication year. For example, MNCS = 2 means that the papers of a university have been cited twice as often as the average of their field and publication year.

PP(top 1%), PP(top 10%), PP(top 50%): the proportion of a university's papers that, compared with other papers in the same field and year, belong to the top 1%, 10%, or 50%, respectively, most frequently cited.

PP(collab): the proportion of a university's papers that have been co-authored with one or more other organizations.

PP(int collab): the proportion of a university's papers that have been co-authored in two or more countries.

For the calculation of the benchmarks for each indicator, we computed the pairwise difference between all universities. In the Leiden Ranking, we have 903 universities with MNCs. The number of pairwise comparisons can be calculated with the formula

$$\frac{x(x-1)}{2}$$

With x = 903, the number of comparisons amounts to 407,253. The mean value of the absolute differences between all pairwise comparisons is proposed here as the benchmark (or expected value) for the interpretation of MNCS differences between two universities. The use of the mean difference as benchmark is only meaningful for size-independent indicators. Thus, we do not consider size-dependent indicators in the calculations which are also available in the Leiden Ranking from the analyses (e.g. *Total Normalized Citation Score* as a variant of the MNCS).

Table 1. Mean difference (M), standard deviation (SD), and maximum value (MAX) for several size-independent indicators in the Leiden Ranking and time periods

| stats | MNCS | PP(top 1%) | PP(top 10%) | PP(top 50%) | PP(collab) | PP(int collab) |
|---|---|---|---|---|---|---|
| 2012-2015 | | | | | | |
| M | 0.34 | 0.01 | 0.04 | 0.08 | 0.08 | 0.16 |
| SD | 0.25 | 0.01 | 0.03 | 0.06 | 0.07 | 0.12 |
| MAX | 1.66 | 0.05 | 0.25 | 0.40 | 0.57 | 0.81 |
| 2011-2014 | | | | | | |
| M | 0.33 | 0.01 | 0.04 | 0.08 | 0.08 | 0.16 |
| SD | 0.25 | 0.01 | 0.03 | 0.06 | 0.07 | 0.12 |
| MAX | 1.66 | 0.05 | 0.25 | 0.41 | 0.61 | 0.80 |
| 2010-2013 | | | | | | |
| M | 0.33 | 0.01 | 0.04 | 0.09 | 0.08 | 0.16 |
| SD | 0.25 | 0.01 | 0.03 | 0.06 | 0.07 | 0.11 |
| MAX | 1.71 | 0.05 | 0.25 | 0.41 | 0.64 | 0.76 |
| 2009-2012 | | | | | | |
| M | 0.33 | 0.01 | 0.04 | 0.09 | 0.09 | 0.15 |
| SD | 0.25 | 0.01 | 0.03 | 0.06 | 0.07 | 0.11 |
| MAX | 1.75 | 0.05 | 0.25 | 0.43 | 0.68 | 0.72 |
| 2008-2011 | | | | | | |
| M | 0.33 | 0.01 | 0.04 | 0.09 | 0.09 | 0.15 |
| SD | 0.25 | 0.01 | 0.03 | 0.07 | 0.07 | 0.11 |
| MAX | 1.59 | 0.05 | 0.24 | 0.42 | 0.68 | 0.76 |
| 2007-2010 | | | | | | |
| M | 0.33 | 0.01 | 0.04 | 0.09 | 0.09 | 0.15 |
| SD | 0.25 | 0.01 | 0.03 | 0.07 | 0.08 | 0.11 |
| MAX | 1.57 | 0.05 | 0.23 | 0.41 | 0.67 | 0.82 |
| 2006-2009 | | | | | | |
| M | 0.33 | 0.01 | 0.04 | 0.09 | 0.09 | 0.15 |
| SD | 0.25 | 0.01 | 0.03 | 0.07 | 0.08 | 0.11 |
| MAX | 1.60 | 0.04 | 0.23 | 0.43 | 0.73 | 0.91 |

The results of the mean differences calculations are presented in Table 1. The table shows the mean of the differences (as well as SD and MAX) between the universities in the Leiden Ranking for several time periods and indicators considered. For example, the mean of the differences between the universities in terms of MNCS is 0.34 (in 2012-2015). This value can be used as a benchmark for assessing the difference between the MNCS of two selected universities. If the difference is larger than the mean value, it is more than one can expect as the difference between two universities. If the difference between the two universities is close to the maximum value in Table 1, it is a remarkable difference. As the key values (mean, sd, and max) in the table show, variation over the time periods is generally low. It seems that the benchmarks are more or less stable. For example, a difference of 4 percentage points in PP (top 10%) separates meaningful from meaningless institutional differences.

The results of Bornmann et al. (2013) reveal that the differences between the universities in the Leiden Ranking are unequally distributed if the universities are ranked according to PP(top 10%). For example, the best performing universities are characterized by larger pairwise differences than universities positioned lower. Thus, one can expect that the mean values presented in Table 1 are different in various performance groups of universities. We tested this assumption by classifying the universities using the Characteristics Scores and Scales (CSS) method – "a parameter-free solution for the assessment of outstanding performance" (Glänzel, Debackere, & Thijs, 2016). Here, we followed the approach of Bornmann and Glänzel (2017). Glänzel and Schubert (1988) introduced the CSS method for assigning the publications in a field and publication year to meaningful impact groups. Characteristic scores are obtained by iteratively truncating samples at their mean and recalculating the mean of the truncated sample until the procedure is stopped or no new scores are generated.

In the first step, we calculated the mean MNCSs for all universities (using "All Science" and "2012-2015" as examples). We classified all universities with a MNCS below the mean as "poorly cited". The universities with MNCS above the mean were used for further calculations in the second step. For these universities the mean MNCS was calculated again and the universities with MNCSs below the mean were assigned to the category "fairly cited". In the third step, we repeated the procedure of mean calculation and separation of two groups which resulted in two further MNCS groups labelled "remarkably cited" and "outstandingly cited".

The results for the mean MNCS differences within several CSS categories on the basis of in-group comparisons of the 903 universities are shown in Table 2. All mean differences are lower than the overall mean difference (with 0.34, see Table 1). This is due to the fact that the universities within the groups are more homogeneous than in the complete set of universities. This result is not unexpected, since the deviation of scores of universities across different groups is expected to be more pronounced then within the same group. The results in Table 2 further reveal that the differences between the universities are larger in the lowest and highest performance groups than in the groups in-between.

Table 2. Mean differences (M), standard deviations (SD), and maximum (MAX) values for MNCS in the Leiden Ranking separated by CSS categories

| CSS category | N universities with N(N–1)/2 comparisons | M | SD | MAX |
| --- | --- | --- | --- | --- |
| Poorly cited | 468 universities with 109,278 comparisons | 0.15 | 0.11 | 0.57 |
| Fairly cited | 248 universities with 30,628 comparisons | 0.09 | 0.06 | 0.26 |
| Remarkably cited | 118 universities with 6903 comparisons | 0.06 | 0.04 | 0.18 |
| Outstandingly cited | 69 universities with 2346 comparisons | 0.15 | 0.13 | 0.65 |

In this Letter to the Editor, we have presented some benchmarks for supporting the interpretation of indicator differences between two universities in the Leiden Ranking. With the following Stata commands, the reader of this Letter is able to produce their own benchmarks for single fields etc. based on the Leiden Ranking data.

**Appendix**
**Stata commands for producing the results in Table 1 ("All sciences", "2006-2009")**

```
**Data import
import excel "CWTS Leiden Ranking 2017.xlsx", sheet("Results") firstrow clear
destring MNCS PP_top1 PP_top10 PP_top50 PP_collab PP_int_collab, replace force
keep if Field == "All sciences"
keep if Period == "2006-2009"
keep if Frac_counting == 0
keep University MNCS PP_top1 PP_top10 PP_top50 PP_collab PP_int_collab

*Analysis of all pairwise combinations
gen id = _n
gen gfreq = _N
expand gfreq
by id, sort: gen numid2 = _n
foreach var of varlist id University MNCS PP_top1 PP_top10 PP_top50 PP_collab PP_int_collab {
gen `var'2 = `var'[gfreq * numid2]
}
drop if id == id2
drop if id > id2
foreach var of varlist MNCS PP_top1 PP_top10 PP_top50 PP_collab PP_int_collab {
gen M_`var'=abs(`var'-`var'2)
```

```
}
```

*Producing the results
tabstat M_MNCS M_PP_top1 M_PP_top10 M_PP_top50 M_PP_collab M_PP_int_collab, stats(n, mean, sd, max) format(%9.2f)